\magnification 1200

%
%
\newdimen\FigSize       \FigSize=.9\hsize 
%
\newskip\abovefigskip   \newskip\belowfigskip
\gdef\epsfig#1;#2;{\par\vskip\abovefigskip\penalty -500
   {\everypar={}\epsfxsize=#1\nd
    \centerline{\epsfbox{#2}}}%
    \vskip\belowfigskip}%
%
\newskip\figtitleskip
\gdef\tepsfig#1;#2;#3{\par\vskip\abovefigskip\penalty -500
   {\everypar={}\epsfxsize=#1\nd
    \vbox
      {\centerline{\epsfbox{#2}}\vskip\figtitleskip
       \centerline{\figtitlefont#3}}}%
    \vskip\belowfigskip}%
%
\newcount\FigNr \global\FigNr=0
\gdef\nepsfig#1;#2;#3{\global\advance\FigNr by 1
   \tepsfig#1;#2;{Figure\space\the\FigNr.\space#3}}%
%
%
%
\gdef\ipsfig#1;#2;{
   \midinsert{\everypar={}\epsfxsize=#1\nd
              \centerline{\epsfbox{#2}}}%
   \endinsert}%
%
\gdef\tipsfig#1;#2;#3{\midinsert
   {\everypar={}\epsfxsize=#1\nd
    \vbox{\centerline{\epsfbox{#2}}%
          \vskip\figtitleskip
          \centerline{\figtitlefont#3}}}\endinsert}%
%
\gdef\nipsfig#1;#2;#3{\global\advance\FigNr by1%
  \tipsfig#1;#2;{Figure\space\the\FigNr.\space#3}}%
\newread\epsffilein    
\newif\ifepsffileok    
\newif\ifepsfbbfound   
\newif\ifepsfverbose   
\newdimen\epsfxsize    
\newdimen\epsfysize    
\newdimen\epsftsize    
\newdimen\epsfrsize    
\newdimen\epsftmp      
\newdimen\pspoints     
\pspoints=1bp          
\epsfxsize=0pt         
\epsfysize=0pt         
\def\epsfbox#1{\global\def\epsfllx{72}\global\def\epsflly{72}%
   \global\def\epsfurx{540}\global\def\epsfury{720}%
   \def\lbracket{[}\def\testit{#1}\ifx\testit\lbracket
   \let\next=\epsfgetlitbb\else\let\next=\epsfnormal\fi\next{#1}}%
\def\epsfgetlitbb#1#2 #3 #4 #5]#6{\epsfgrab #2 #3 #4 #5 .\\%
   \epsfsetgraph{#6}}%
\def\epsfnormal#1{\epsfgetbb{#1}\epsfsetgraph{#1}}%
\def\epsfgetbb#1{%
%
%
\openin\epsffilein=#1
\ifeof\epsffilein\errmessage{I couldn't open #1, will ignore it}\else
%
%
   {\epsffileoktrue \chardef\other=12
    \def\do##1{\catcode`##1=\other}\dospecials \catcode`\ =10
    \loop
       \read\epsffilein to \epsffileline
       \ifeof\epsffilein\epsffileokfalse\else
%
%
          \expandafter\epsfaux\epsffileline:. \\%
       \fi
   \ifepsffileok\repeat
   \ifepsfbbfound\else
    \ifepsfverbose\message{No bounding box comment in #1; using
defaults}\fi\fi
   }\closein\epsffilein\fi}%
%
%
\def\epsfsetgraph#1{%
   \epsfrsize=\epsfury\pspoints
   \advance\epsfrsize by-\epsflly\pspoints
   \epsftsize=\epsfurx\pspoints
   \advance\epsftsize by-\epsfllx\pspoints
%
%
   \epsfxsize\epsfsize\epsftsize\epsfrsize
   \ifnum\epsfxsize=0 \ifnum\epsfysize=0
      \epsfxsize=\epsftsize \epsfysize=\epsfrsize
%
%
     \else\epsftmp=\epsftsize \divide\epsftmp\epsfrsize
       \epsfxsize=\epsfysize \multiply\epsfxsize\epsftmp
       \multiply\epsftmp\epsfrsize \advance\epsftsize-\epsftmp
       \epsftmp=\epsfysize
       \loop \advance\epsftsize\epsftsize \divide\epsftmp 2
       \ifnum\epsftmp>0
          \ifnum\epsftsize<\epsfrsize\else
             \advance\epsftsize-\epsfrsize \advance\epsfxsize\epsftmp
\fi
       \repeat
     \fi
   \else\epsftmp=\epsfrsize \divide\epsftmp\epsftsize
     \epsfysize=\epsfxsize \multiply\epsfysize\epsftmp
     \multiply\epsftmp\epsftsize \advance\epsfrsize-\epsftmp
     \epsftmp=\epsfxsize
     \loop \advance\epsfrsize\epsfrsize \divide\epsftmp 2
     \ifnum\epsftmp>0
        \ifnum\epsfrsize<\epsftsize\else
           \advance\epsfrsize-\epsftsize \advance\epsfysize\epsftmp \fi
     \repeat
   \fi
%
%
   \ifepsfverbose\message{#1: width=\the\epsfxsize,
height=\the\epsfysize}\fi
   \epsftmp=10\epsfxsize \divide\epsftmp\pspoints
   \vbox to\epsfysize{\vfil\hbox to\epsfxsize{%
      \includegraphics{#1}%
      \hfil}}%
\epsfxsize=0pt\epsfysize=0pt}%
%
%
{\catcode`\%=12
\global\let\epsfpercent=
%
%
\long\def\epsfaux#1#2:#3\\{\ifx#1\epsfpercent
   \def\testit{#2}\ifx\testit\epsfbblit
      \epsfgrab #3 . . . \\%
      \epsffileokfalse
      \global\epsfbbfoundtrue
   \fi\else\ifx#1\par\else\epsffileokfalse\fi\fi}%
%
%
\def\epsfgrab #1 #2 #3 #4 #5\\{%
   \global\def\epsfllx{#1}\ifx\epsfllx\empty
      \epsfgrab #2 #3 #4 #5 .\\\else
   \global\def\epsflly{#2}%
   \global\def\epsfurx{#3}\global\def\epsfury{#4}\fi}%
%
%
\def\epsfsize#1#2{\epsfxsize}%
%
%

\epsfverbosetrue                        
\abovefigskip=\baselineskip             
\belowfigskip=\baselineskip             
\global\let\figtitlefont\bf             
\global\figtitleskip=.5\baselineskip    

\font\tenmsb=msbm10   
\font\sevenmsb=msbm7
\font\fivemsb=msbm5
\newfam\msbfam
\textfont\msbfam=\tenmsb
\scriptfont\msbfam=\sevenmsb
\scriptscriptfont\msbfam=\fivemsb

\let\nd\noindent 

\def\natural{{\rm I\kern-.18em N}}
\newskip\ttglue


\def\eightpoint{\def\rm{\fam0\eightrm}  
  \textfont0=\eightrm \scriptfont0=\sixrm \scriptscriptfont0=\fiverm
  \textfont1=\eighti  \scriptfont1=\sixi  \scriptscriptfont1=\fivei
  \textfont2=\eightsy  \scriptfont2=\sixsy  \scriptscriptfont2=\fivesy
  \textfont3=\tenex  \scriptfont3=\tenex  \scriptscriptfont3=\tenex
  \textfont\itfam=\eightit  \def\it{\fam\itfam\eightit}
  \textfont\slfam=\eightsl  \def\sl{\fam\slfam\eightsl}
  \textfont\ttfam=\eighttt  \def\tt{\fam\ttfam\eighttt}
  \textfont\bffam=\eightbf  \scriptfont\bffam=\sixbf
    \scriptscriptfont\bffam=\fivebf  \def\bf{\fam\bffam\eightbf}
  \tt  \ttglue=.5em plus.25em minus.15em
  \normalbaselineskip=9pt
  \setbox\strutbox=\hbox{\vrule height7pt depth2pt width0pt}
  \let\sc=\sixrm  \let\big=\eightbig \normalbaselines\rm}

\font\eightrm=cmr8 \font\sixrm=cmr6 \font\fiverm=cmr5
\font\eighti=cmmi8  \font\sixi=cmmi6   \font\fivei=cmmi5
\font\eightsy=cmsy8  \font\sixsy=cmsy6 \font\fivesy=cmsy5
\font\eightit=cmti8  \font\eightsl=cmsl8  \font\eighttt=cmtt8
\font\eightbf=cmbx8  \font\sixbf=cmbx6 \font\fivebf=cmbx5

\def\eightbig#1{{\hbox{$\textfont0=\ninerm\textfont2=\ninesy
        \left#1\vbox to6.5pt{}\right.\enspace$}}}

\def\ninepoint{\def\rm{\fam0\ninerm}  
  \textfont0=\ninerm \scriptfont0=\sixrm \scriptscriptfont0=\fiverm
  \textfont1=\ninei  \scriptfont1=\sixi  \scriptscriptfont1=\fivei
  \textfont2=\ninesy  \scriptfont2=\sixsy  \scriptscriptfont2=\fivesy
  \textfont3=\tenex  \scriptfont3=\tenex  \scriptscriptfont3=\tenex
  \textfont\itfam=\nineit  \def\it{\fam\itfam\nineit}
  \textfont\slfam=\ninesl  \def\sl{\fam\slfam\ninesl}
  \textfont\ttfam=\ninett  \def\tt{\fam\ttfam\ninett}
  \textfont\bffam=\ninebf  \scriptfont\bffam=\sixbf
    \scriptscriptfont\bffam=\fivebf  \def\bf{\fam\bffam\ninebf}
  \tt  \ttglue=.5em plus.25em minus.15em
  \normalbaselineskip=11pt
  \setbox\strutbox=\hbox{\vrule height8pt depth3pt width0pt}
  \let\sc=\sevenrm  \let\big=\ninebig \normalbaselines\rm}

\font\ninerm=cmr9 \font\sixrm=cmr6 \font\fiverm=cmr5
\font\ninei=cmmi9  \font\sixi=cmmi6   \font\fivei=cmmi5
\font\ninesy=cmsy9  \font\sixsy=cmsy6 \font\fivesy=cmsy5
\font\nineit=cmti9  \font\ninesl=cmsl9  \font\ninett=cmtt9
\font\ninebf=cmbx9  \font\sixbf=cmbx6 \font\fivebf=cmbx5
\def\ninebig#1{{\hbox{$\textfont0=\tenrm\textfont2=\tensy
        \left#1\vbox to7.25pt{}\right.$}}}

\def\chix{{\raise.5ex\hbox{$\chi$}}}
\def\chixa{{\chix\lower.2em\hbox{$_A$}}}

\def\real{{\rm I\kern-.2em R}}
\def\integer{{\rm Z\kern-.32em Z}}
\def\complex{\kern.1em{\raise.47ex\hbox{
            $\scriptscriptstyle |$}}\kern-.40em{\rm C}}
\def\vs#1 {\vskip#1truein}
\def\hs#1 {\hskip#1truein}
  \hsize=6.2truein \hoffset=.23truein 
  \vsize=8.8truein 
\pageno=1 \baselineskip=12pt
  \parskip=0 pt \parindent=20pt 
\overfullrule=0pt \lineskip=0pt \lineskiplimit=0pt
  \hbadness=10000 \vbadness=10000 
     \pageno=0
     
     \footline{\ifnum\pageno=0\hss\else\hss\tenrm\folio\hss\fi}
     \hbox{}
     \vskip 1truein\centerline{{\bf Dilatancy Transition in a Granular Model}}
     \vskip .2truein\centerline{by}
     \vskip .2truein
\centerline{{David Aristoff}
\ \ and\ \  {Charles Radin}
\footnote{*}{Research supported in part by NSF Grant DMS-0700120\hfil}}

\vskip .1truein
\centerline{ Mathematics Department, University of Texas, Austin, TX 78712} 
\vs.5 \centerline{{\bf Abstract}} 

\vs.1 \nd We introduce a model of
     granular matter and use a stress ensemble to analyze shearing.
  Monte Carlo simulation shows the model to exhibit
     a second order phase transition, associated
     with the onset of dilatancy.

\vs3
\centerline{May, 2010}
\vs1
\centerline{PACS Classification:\ \ 45.70.Cc, 81.05.Rm, 05.70.Ce}
     \vfill\eject
\nd
{\bf 1. Introduction.}
\vs.1

Static granular matter, such as a sand pile, can exist in a range of
densities. In its densely packed state, the more common form, it
expands under shear, while when  its grains are loosely packed it
contracts under shear [RN]. We introduce a model within which 
the transition between these qualitative behaviors is singular in the
precise traditional sense of a second order phase transition. (For
two dimensional treatments see [AS, PLR], and for an interpretation as
a glass transition see [CH].)

Our ``granular hard cubes'' model is a granular variant of the classical hard
cubes model of equilibrium statistical mechanics [HHB], the latter being a simplification
of the classical hard sphere model in which the spheres are replaced
by nonoverlapping, parallel unit cubes. In our granular version,
following Edwards [EO], we only allow configurations in which the
parallel cubes are mechanically stable under gravity, where gravity is
acting parallel to edges of the cubes. We require that the centers of
our cubes be approximately at the vertices of an fcc lattice, which
we orient so that gravity acts along the $001$ axis, so the fcc lattice
is thought of as parallel 2-dimensional square lattice layers. We enforce this
approximate structure by requiring that each cube sit on exactly four cubes in
the layer below
it; see Fig.\ 1.
\vs0
\epsfig .6\hsize; 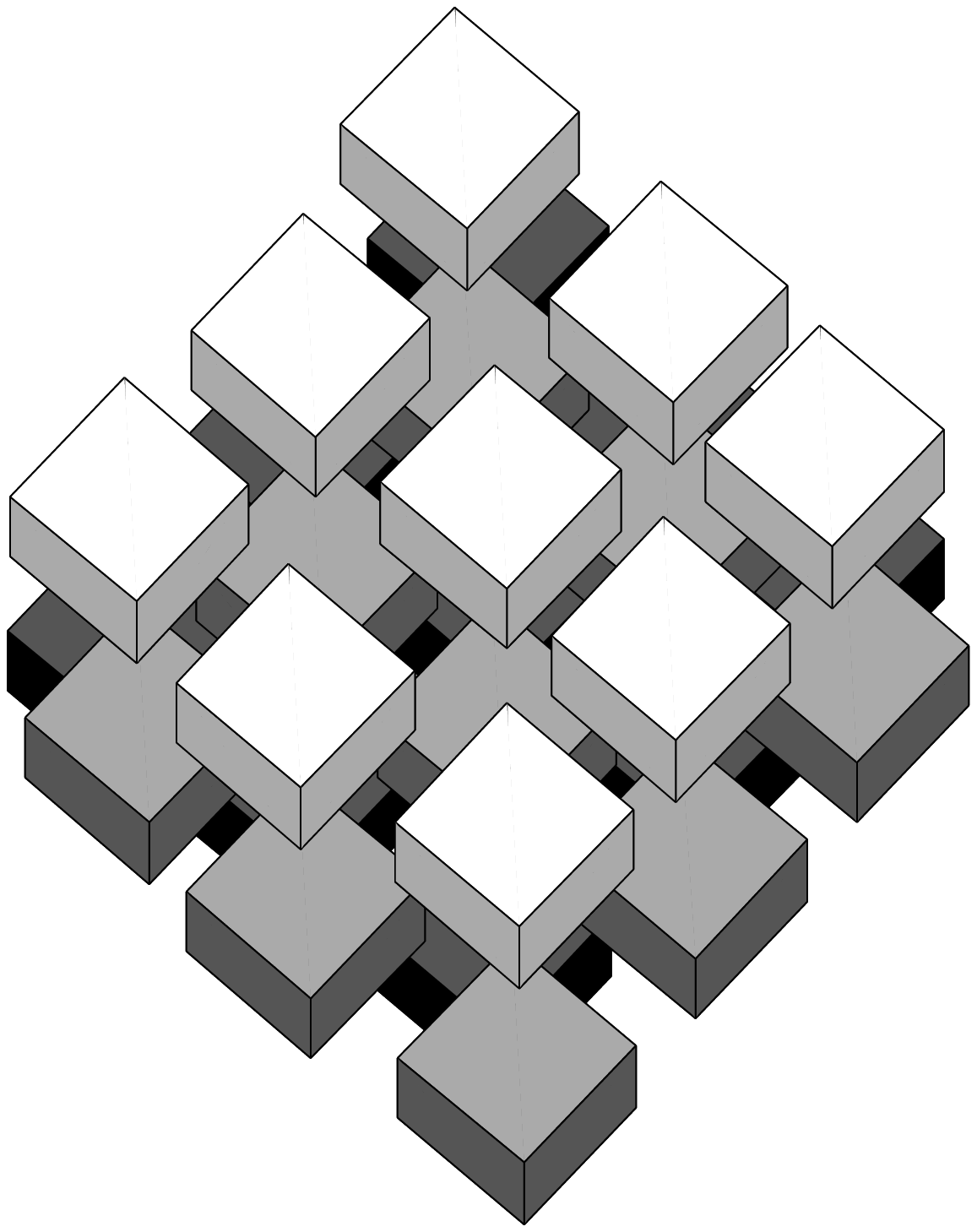;
\vs-.3
\centerline{Figure 1.\ Layers of parallel cubes.}
\vs.2

We want to model the reaction of a sandpile to shear and in particular
we need a model which exhibits a critical state density, a density
separating a high density regime of states, in which the material
expands under shear, from a low density regime in which it contracts
under shear [RN]. As we show, this is easy to arrange. The more interesting
question is whether the model predicts that the transition through
that critical state density is smooth, or is singular.

To understand why it is natural for a granular model 
to have a critical state density it is
useful to use the following ``stress'' ensemble. Instead of constant
density we introduce a pressure parameter, and in place of constant
strain we introduce a shear stress parameter. In other words the states of
the model, which, in the general approach of Edwards [EO] 
are probability densities of configurations of grains, 
optimize the free energy 
$$F(p,f) := S - pV + f\alpha V,\eqno{(1)}$$
where $S=\ln(Z_{p,f})$ is the entropy, 
$V$ is the volume in physical space, and $\alpha$ is the shear strain. 
The parameter $p$ is pressure, $f$ is shear stress, and 
$$Z_{p,f}=\int \int \exp(-pV + f\alpha V)\,dVd\alpha.\eqno{(2)}$$
\nd This is equivalent to saying that a configuration $C$ has an
unnormalized probability density:
$$Pr(C) = \exp(-pV + f\alpha V).\eqno{(3)}$$

It follows from this probability density that, in the absence of shear,
the state of lowest possible particle density, called random loose packing,
occurs at zero pressure [AR1] and some shear strain $\alpha = \alpha_0$. 
Therefore at sufficiently low pressure if
one shears the system from $\alpha_0$, 
the density has no way to change except to 
increase. So at sufficiently low density there is a regime 
in which the system contracts.

On the other hand, at least since Reynolds [Re] one understands the expansion of (dense)
granular matter under shear, which he termed dilatancy, in geometric terms, as
the need for parallel layers of spheres to get out of each others' way
under a strain deforming the layers, either by separating and/or by
thinning within the layers. So we adjust the model to arrange for this
(thinning) phenomenon, as follows.
\vs.1
\epsfig .7\hsize; 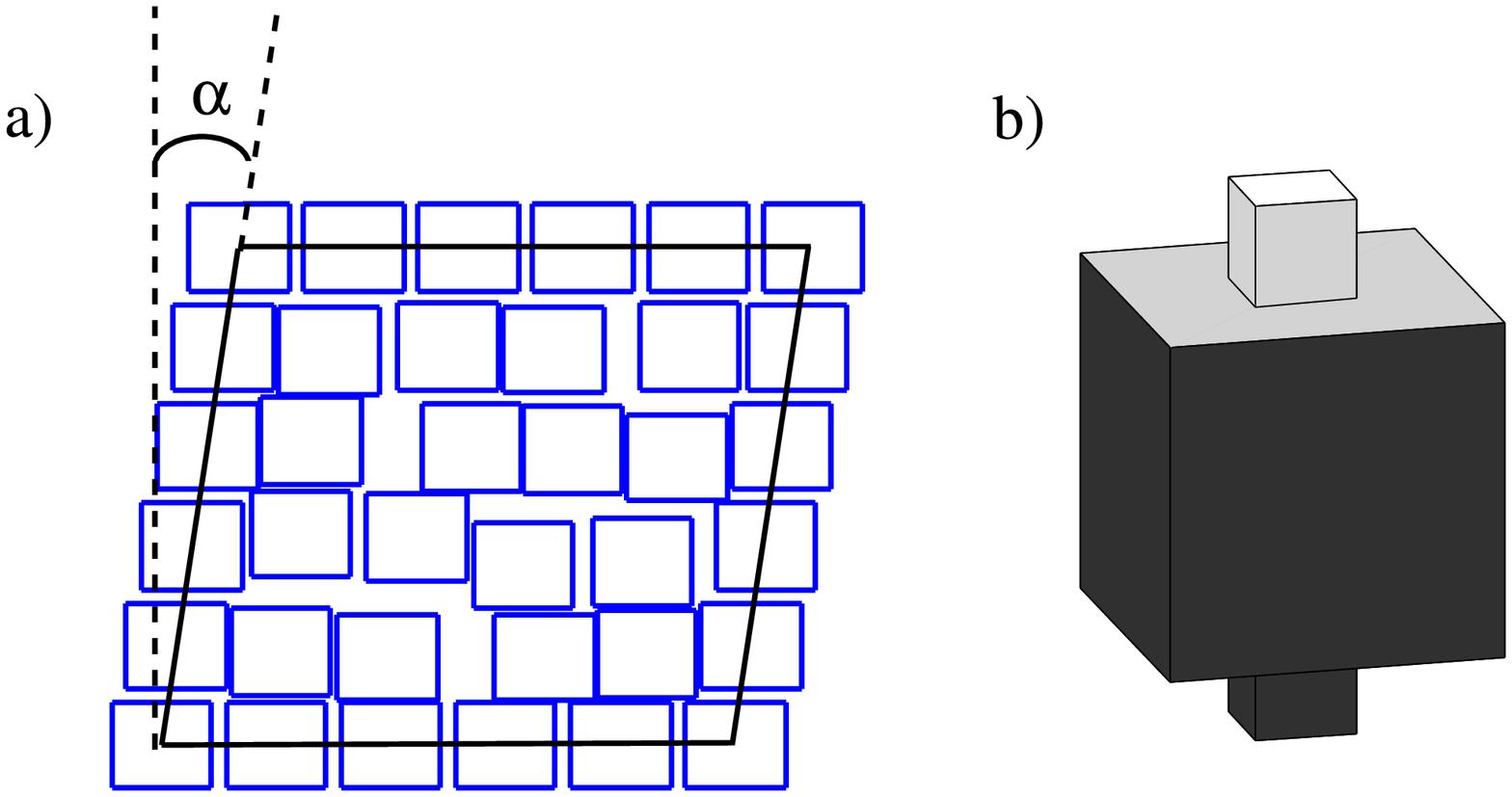;
\vs0
\nd {Figure 2.\ a) A view from above of a layer in a configuration, with a definition}
{of the angle $\alpha$;\  b) A cubic grain, with bumps on top and bottom faces.}
\vs.1 
To react to the strain (see Fig.\ 2a) we mimic the spherical caps that grains in one
layer present to grains in neighboring layers by adding ``bumps'' to the
top and bottom of our
cubes; see Fig.\ 2b. Such cubes will, when we
strain a sufficiently dense configuration in response to the shear, feel the bumps in
the layers above and below it, and this will produce a dilatancy effect by thinning
the layers. By considering optimally symmetric configurations one quickly 
finds $\phi_d := (1+w)^{-1}$, where $w$ is the width of a bump, as
an estimate of the density required for dilatancy; see Fig.\ 3. 
In our simulations we use $w = 0.15$, and so $\phi_d \approx 0.59$. The 
qualitative dilatancy effect is not sensitive to the precise value of
$w$, though the value $\phi_d$ is.

So the bumps,
together with the universal behavior at low pressure or density, explain the
existence of a critical state density. The main question then
is: as density is varied, does the transition through the
critical state density proceed in a smooth or in a singular manner? We
will show there is an unambiguous second order phase transition at
dilatancy onset, at density near $\phi_d$. This suggests, by analogy
with matter in thermal equilibrium,
that the material in the two regimes differs in other
characteristics as well, for instance it would be expected that the
yield force behave differently in the two regimes. In [SNRS] the yield force is measured
as a function of density, and a second order phase transition is
found at a density roughly $0.598$. The critical state density was not carefully
measured because of the experimental setup, but approximately
coincided. Our result suggests the two phenomena are simply different
manifestations of the same transition.

We also note that the behavior of granular matter at the random close
packing density, about  $0.64$,
has been interpreted in [Ra, AR2] as a first order phase transition in
which the high density regime is an ordered phase. Together with our
results here, this means that the
usual freezing transition of equilibrium fluids, at which the material
acquires the solid features of an ordered (typically crystalline)
internal structure as well as a strong resistance to shear, seems to
be split into two stages in granular materials, with the resistance to
shear occuring at dilatancy onset at density about $0.6$ and the
ordered internal structure occuring at the random close packing density of
about $0.64$.

\vs-.1
\epsfig .4\hsize; 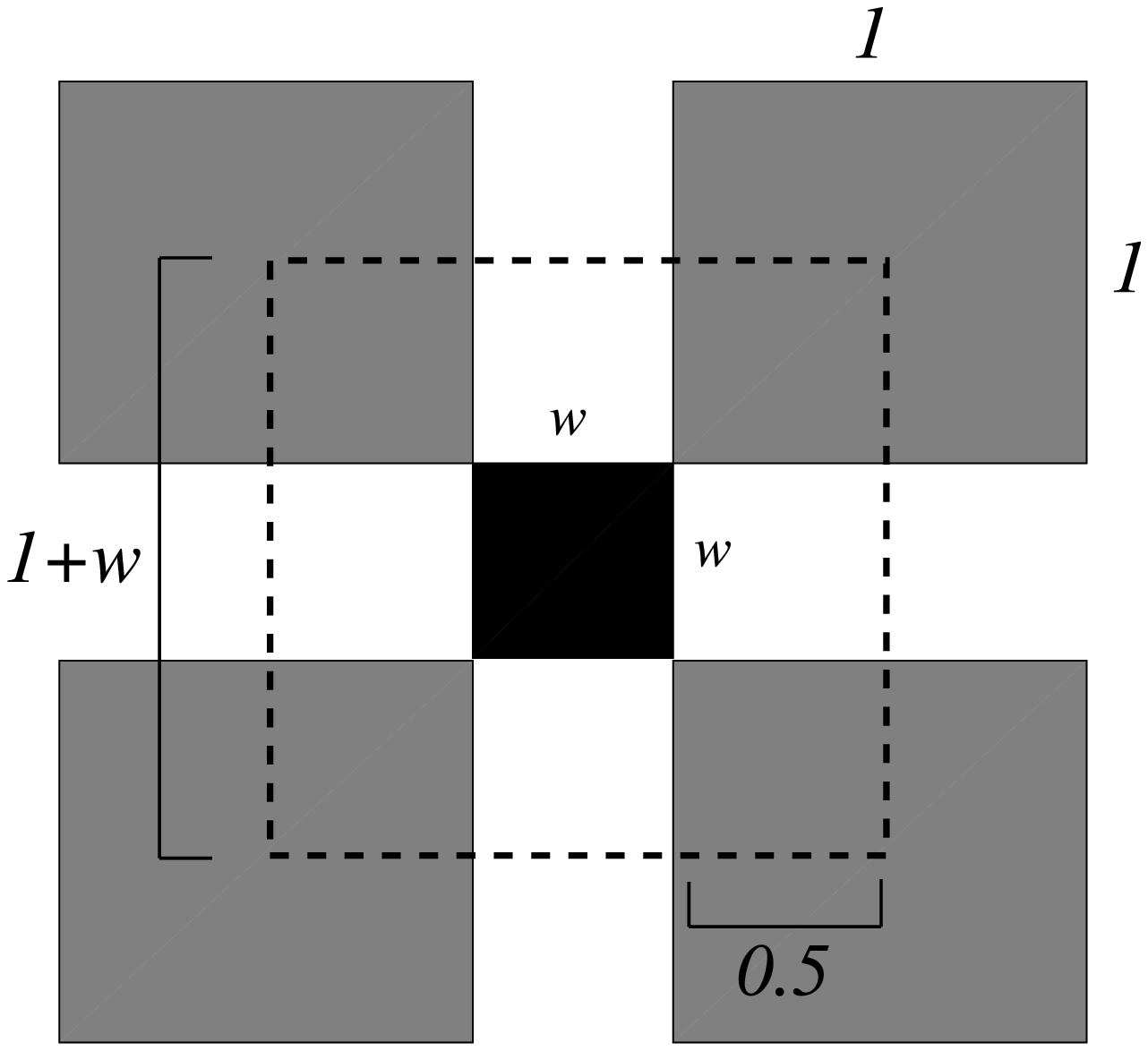;

\vs0
\nd {Figure 3.\ A regular arrangement of grains at maximum density, from above.}
{The configuration can be partitioned into unit cells defined by the dotted lines.}

\vs2
\nd {\bf 2. Simulating the Model.}
\vs.1

As noted above, our model consists of hard, parallel, oriented 
unit cubic grains, with two small cubic bumps of width $w \ll 1$ attached 
to the middle of their upper and lower faces. We use $w = 0.15$. 
Each grain must sit on exactly 
four other grains (except grains on the outside of the configuration), 
but not on the bumps, and grains cannot 
overlap. Thus, the grains appear on discrete vertical levels, so the 
distance in the vertical or $z$-direction between the centers of grains on 
adjacent levels is equal to 1; see Fig.\ 4. 
 
\vs0
\epsfig .6\hsize; 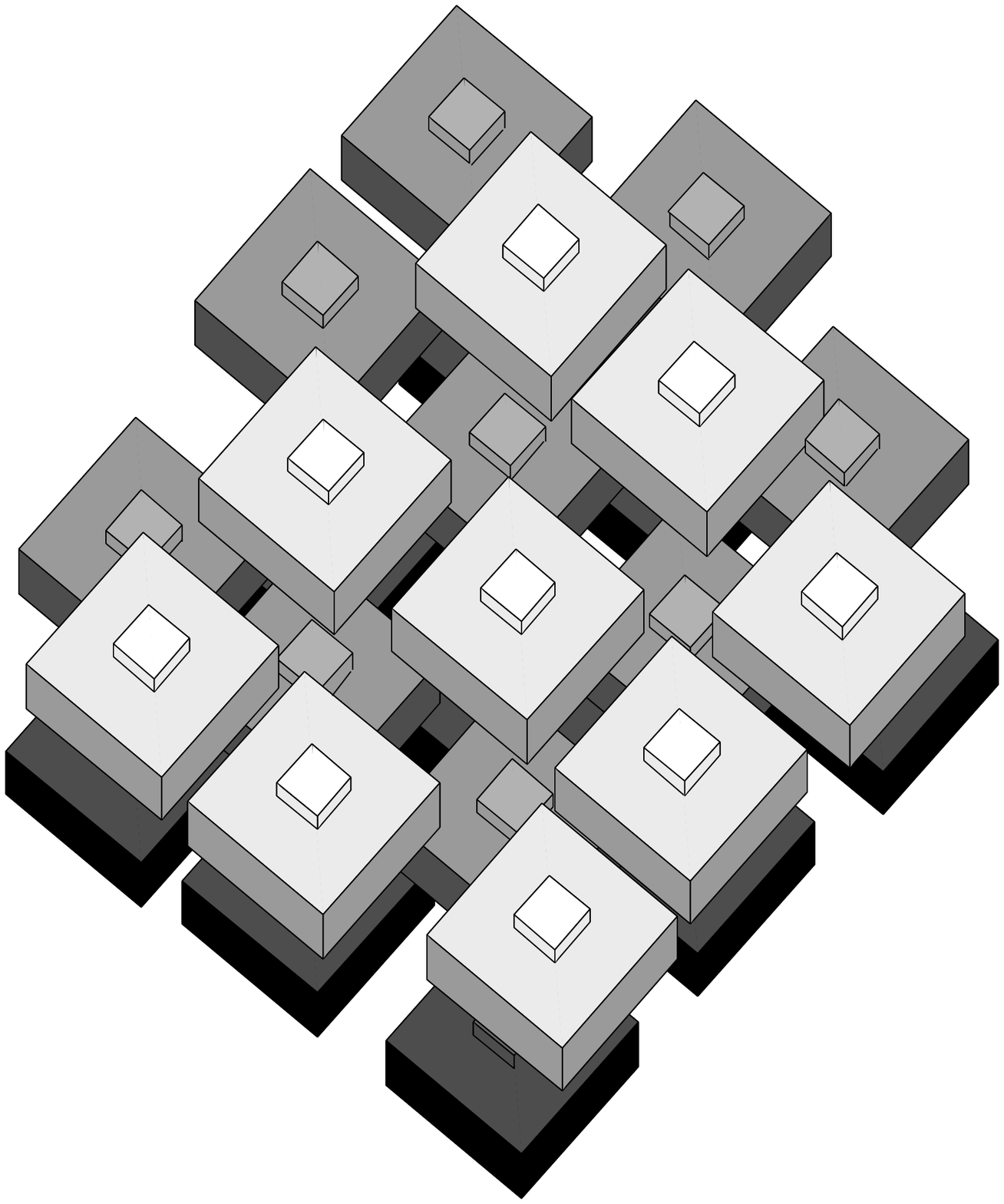;

\vs-.2
\centerline{Figure 4.\ An arrangement of grains, viewed from above.}

\vs.2

Each simulation begins with a perfectly regular arrangement of $N = n^3$ grains in a large 
cubic structure, so that there are $n^2$ grains in each level, with 
$n$ in each row and each column of a level, and $n$ total levels. (For ease in notation 
we associate a row with the $x$-direction and a column with the $y$-direction; 
note that the model has a well-defined row and column structure.) 
The grains on the side boundaries of the cubic structure are allowed to 
make coordinated moves corresponding to changes in configuration 
volume and shape, but they are not allowed to move singly.

We allow three types of Monte Carlo steps. 
With probability $1-1/N$ we choose a random grain inside the boundary 
to move in either the $x$- or $y$-direction; and with probability $1/N$
we allow a configuration to change shape 
or volume. When a configuration changes volume, the $x$- and $y$-coordinates of 
each grain, on each level, are scaled by the same factor $\lambda$. 
When it changes shape, the $x$-coordinate of each grain in 
the $k$th column of each level is scaled, for some $\alpha$, by the factor 
$\alpha k$.

To prevent grains from getting stuck between the 
boundary grains, we impose top-bottom 
periodic boundary conditions, and we remove the bumps from 
the grains nearest to the boundary and on the boundary.

As is standard in Monte Carlo, to simulate the distribution (3)
we choose steps with relative acceptance probabilities 
given by $V^N \exp(-pV + f\alpha V)$. (The factor $V^N$ arises 
from the necessity of rescaling position coordinates to live in 
a fixed-size box [Mc]). Here $\alpha$ is the 
angle of deformation of a configuration (see Fig. 2b), where from symmetry
we can assume
$\alpha \ge 0$ without loss of generality. 
The volume $V$ of a configuration is the volume of the convex hull of the 
set of centers of all the grains. The volume can be computed 
directly from the scale factor $\lambda$ described above.

\vs.2 \nd
{\bf 3. Results of the Simulations.}
\vs.1

We ran Markov chain Monte Carlo simulations on the model, with the Monte Carlo 
steps designed so that the stationary distribution of the Markov chain has the 
probability density 
$$m_{p,f}(C) = {\exp(-pV + f\alpha V)\over Z_{p,f}}\eqno{(4)}$$
described above. We want to measure how the (average) density $\phi = N/V$ changes
as shear stress $f$ varies near $f = 0$. (For convenience we do not include the 
bumps in the calculation of $\phi$, but since particle number $N$ is fixed in our 
simulation, the ``true'' volume fraction is just a constant multiple of
$\phi$.) 
That is, we want to estimate the derivative
$$D := {{\partial \langle \phi\rangle }\over{\partial f}}\bigg|_{f=0}\eqno{(5)}$$
of the average of $\phi$ as a function of $p$.  
 From the definition of $m_{p,f}$ it is 
easy to see that
$$D = D(p) = N\langle \alpha\rangle_{p,0} - \Big[\langle \phi\rangle_{p,0}\Big]
\Big[\langle V\alpha\rangle_{p,0}\Big],\eqno{(6)}$$
\nd where $\displaystyle \langle \cdot\rangle_{p,f}$ denotes average
with respect to $m_{p,f}$.
Thus, to estimate $D(p)$ from our simulations we set $f = 0$ and 
calculate the sample averages of $N\alpha$, $\phi$, and $V\alpha$. With the free energy 
$F(p,f) = S - pV + f\alpha V$, which is the quantity minimized by the state $m_{p,f}$, one can see 
from a direct calculation that 
$$D(p) = -\langle \phi\rangle ^2_{p,0}{{\partial}\over{\partial f}}\left({{\partial F}\over{\partial p}}\right)\bigg|_{f=0},\eqno{(7)}$$
which shows that the function $D(p)$ is a quantity for
which a discontinuity might reasonably be described as a 
second order phase transition.

\vs.1
\epsfig .7\hsize; 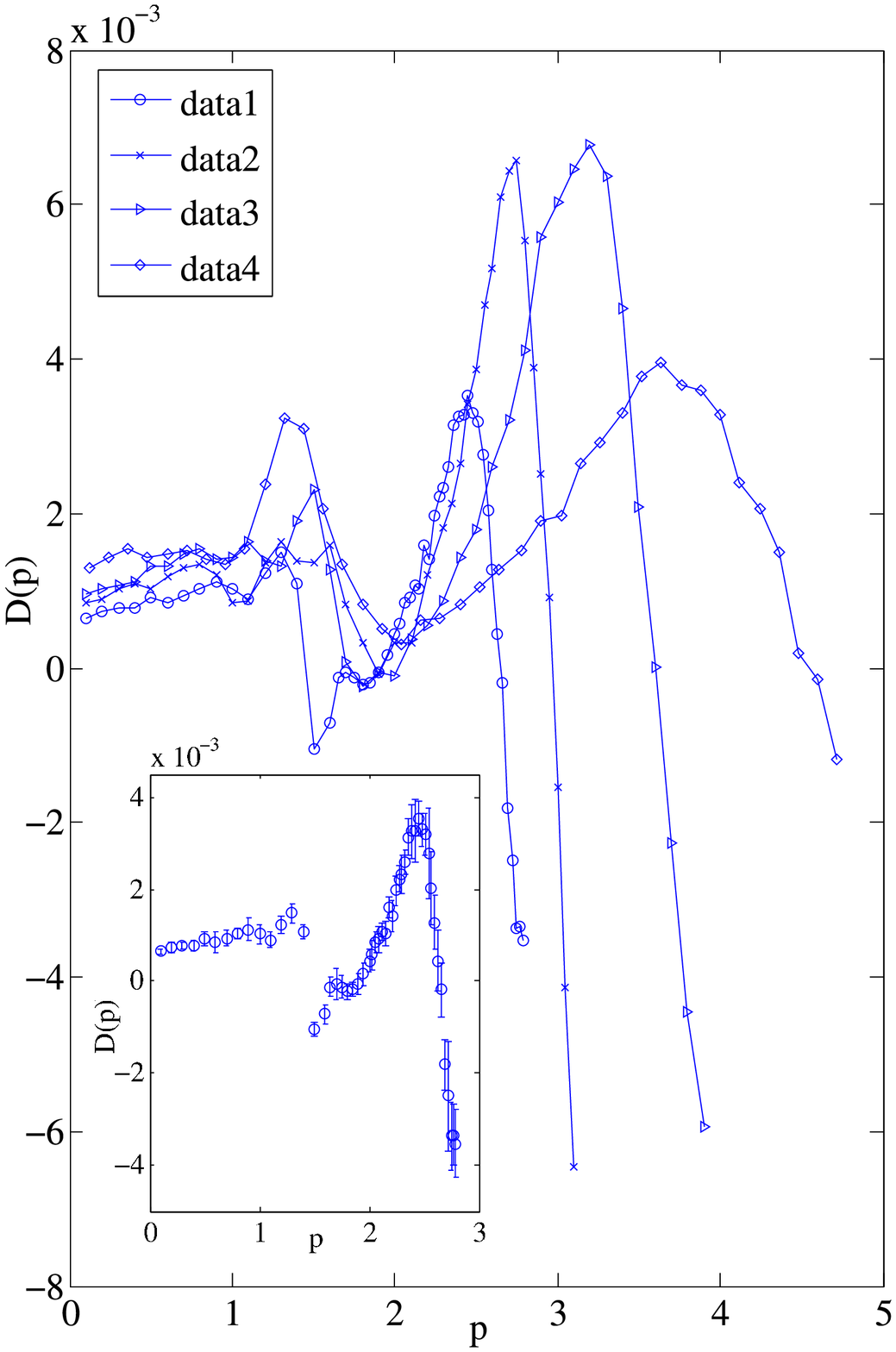;
\vs0
\nd {Figure 5.\  The graph of $\displaystyle D(p) = {{\partial
\langle \phi\rangle_{p,f} }\over {\partial f}}\bigg|_{f=0}$, for systems 
of $14^3=2744$, $12^3=1728$, $10^3=100$, and $8^3=512$ grains (data1-data4, resp.). 
The insert gives error bars on the system with $14^3$ grains.}
\vs.2
\nd

We investigate systems of $8^3=512$, $10^3=100$, $12^3=1728$ and $14^3=2744$ grains at pressures 
$p_1<...<p_k$, where the pressures chosen are different for each size system. In the simulations 
we begin at low pressure $p_1$, and then slowly increase $p$ until just after the 
derivative $D(p)$ falls sharply below zero, roughly corresponding to pressure $p_k$. 
The last configuration in the simulation of $p_i$ is 
used as the starting configuration of the simulation of $p_{i+1}$. We use the standard biased 
autocorrelation function to determine a ``mixing time'', measured as the number of 
Monte Carlo steps required before the 
autocorrelation first crosses zero. We run our simulations long enough so that the simulation 
of each $p_i$ contains, on average, at least $20$ mixing times, and we run $200$ independent 
copies of each simulation to obtain the averages $\langle N\alpha\rangle_{p,0}$, 
$\langle \phi\rangle_{p,0}$, and 
$\langle V\alpha\rangle_{p,0}$. From these averages we compute $D(p)$
from (6). 
(We also ran some of our simulations much longer, with fewer copies 
and pressures, and noted agreement with the already-obtained data on $D(p)$.) 
Then, for error bars on $D(p)$, 
we repeat the entire experiment $8$ times and use the Student's $t$-distribution.

Our simulations suggest that $D(p)$ develops a discontinuity as the system size increases 
(see Fig.\ 5). Changing variables we note that $D$, as a function of $\langle \phi \rangle_{p,0}$, develops 
a discontinuity near the (average) volume fraction $\phi_d \approx 0.59$ discussed in the introduction;
we plot $\langle \phi \rangle_{p,0}$ against $p$ in Fig.\ 6b and $D$ 
against $\langle \phi \rangle_{p,0}$ in Fig.\ 6a.

Note that for $p \le 1$, 
$D(p)$ exhibits regular behavior in which $D(p)$ is roughly constant, $D(p)=\eta \approx 0.001$. 
Then for 
$1 < p < 2$, $D(p)$ has some oscillation which is characteristic to the system size. 
Then for $p \ge 2$, $D(p)$ steadily increases to a peak, then sharply
decreases through zero. We believe that $D(p)$ is discontinuous in the limit of infinite
system size, but the rate of change of $D(p)$ is so large it is difficult to measure its 
variation with system size. Instead we consider two measures of 
the interval $R$ in which the discontinuity is developing, and then 
note that $R$ gets smaller and smaller as system size increases.

We expect that the oscillation observed in $D(p)$ in the interval $1< p < 2$ is caused 
by finite-size effects, so that it disappears in the limit $N \to \infty$. 
Furthermore we expect  
the critical pressure $p_c$ to fall at either the end or the beginning of 
the oscillation region. These two possibilities underlie, respectively, 
our measurements $R_1$ and $R_2$, defined below.
Let $\eta$ be the average value of $D(p)$ over $p \le 1$. 
The left endpoint of $R_1$ is the smallest value of $p$ 
where $p > 2$ and $D(p) \approx \eta$, and the right endpoint of $R_1$
is the value of $p$ where $p > 2$ and  $D(p) \approx 0$. 
The widths of $R_1$ are approximately $1.75\pm0.05$, $1.15\pm0.05$, 
$0.75\pm0.05$, and $0.50\pm0.05$ for systems of 
$N = 8^3$, $10^3$, $12^3$ and $14^3$ grains, respectively.  
On the other hand, the left endpoint of $R_2$ is defined as the 
smallest value of $p$ such that $D(p)$ 
differs significantly from $\eta$, while the right endpoint of $R_2$ 
is the same as the right 
endpoint of $R_1$. We estimate that the widths of
$R_2$ are approximately $3.55\pm0.05$, $1.55\pm0.05$, 
$1.25\pm0.15$, and $1.15\pm0.05$ for systems of 
$N = 8^3$, $10^3$, $12^3$ and $14^3$ grains, respectively. 

In either case, based on the decreasing sizes of the intervals $R_1$ and $R_2$, we believe that the jump 
from $D(p) \approx \nu $ to $D(p) \ll 0$ occurs over an interval which is vanishing in the infinite volume 
limit, which, given the relation between $D(p)$ and the free
energy $F(p,f)$ in (7),
is reasonably termed a second order phase transition. And as noted in
the introduction, there is experimental evidence for this
interpretation in [SNRS].
\vs0
\epsfig 1.0\hsize; 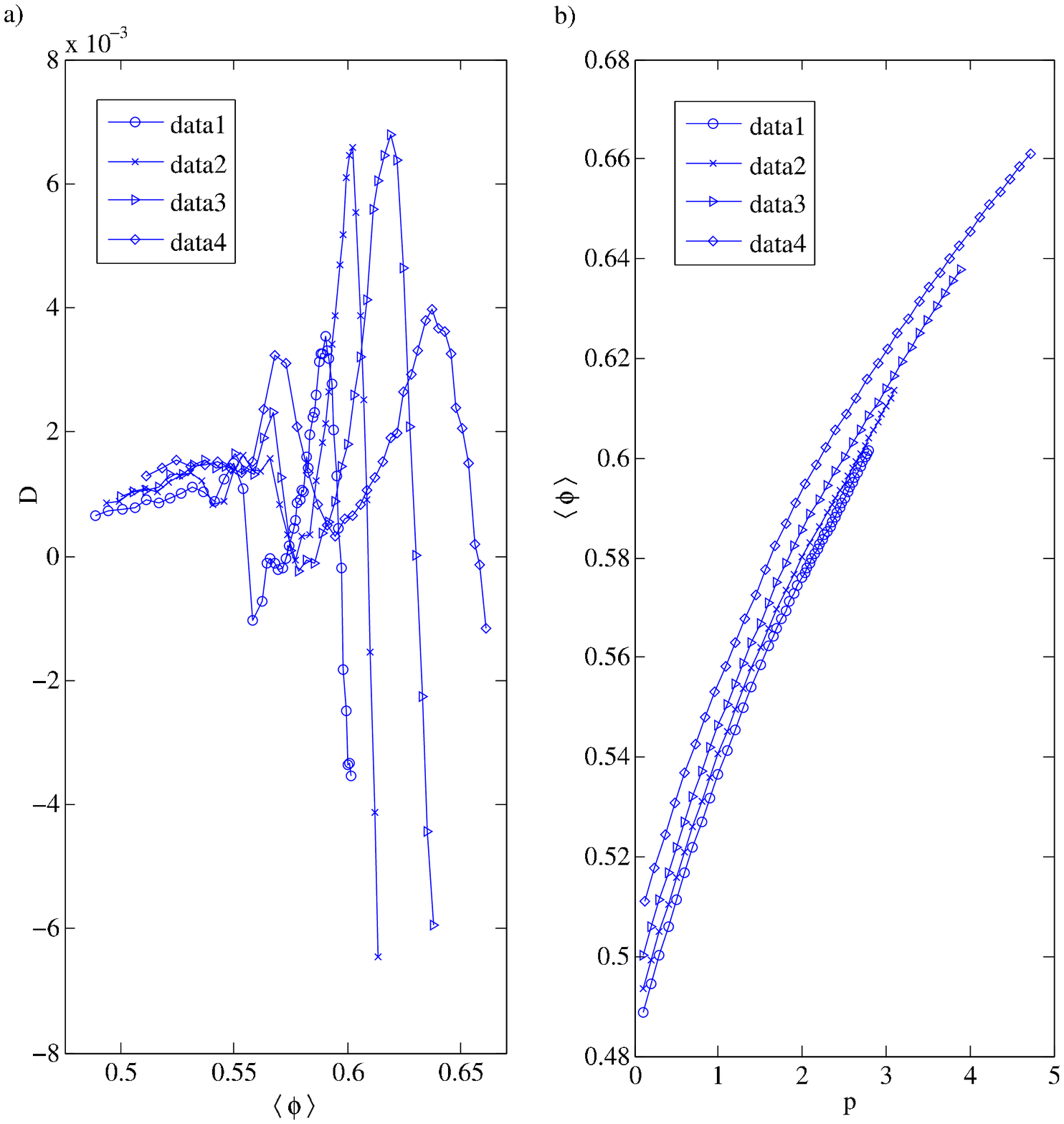;
\vs0
\nd {Figure 6.\ a) $D(p)$ as a function of $\displaystyle \langle
\phi\rangle_{p,0}$; \  b) $\displaystyle \langle
\phi\rangle_{p,0}$ as a function of $p$. In both plots, 
data1-data4 represent systems of $14^3=2744$, $12^3=1728$, 
$10^3=100$, and $8^3=512$ grains, respectively.}
\vs.5
\nd
{\bf Acknowledgements.} It is a pleasure to acknowledge useful
discussions with Hans C.\ Andersen and Matthias Schr\"oter.

\vfill \eject
\vs.2 
\centerline{References}
\vs.2
\item{[AR1]}D.\ Aristoff and C.\ Radin,
Random loose packing in granular matter,
J.\ Stat.\ Phys.\ 135 (2009) 1-23.

\item{[AR2]}D.\ Aristoff and C.\ Radin, 
Random close packing in a granular model,

\hs.03 arXiv:0909.2608.

\item{[AS]}E.\ Aharonov and D.\ Sparks,
Rigidity phase transition in granular packings,
Phys. Rev. E 60 (1999) 6890-6896.

\item{[CH]}A Coniglio and H.J Herrmann,
Phase transitions in granular packings,
Physica A 225 (1996) 1-6.

\item{[HHB]}W.G Hoover, C.G Hoover and M.N.\ Bannerman,
Single-Speed Molecular Dynamics of Hard Parallel Squares and Cubes,
J.\ Stat.\ Phys.\ 136  (2009) 715-732.

\item{[EO]} S.F.\ Edwards and R.B.S.\ Oakeshott, 
Theory of powders, 
Physica A 157 (1989) 1080-1090.

\item{[Mc]} I.R.\ MacDonald,
NpT-ensemble Monte Carlo calculations for binary liquid mixtures, 
Molecular Physics Vol 100 (2002) 95-105.

\item{[PLR]}M.\ Piccioni, V.\ Loreto, and S.\ Roux,
Criticality of the ``critical state'' of granular media: Dilatancy
angle in the Tetris model,
Phys.\ Rev.\ E 61 (2000) 2813-2817.


\item{[Ra]}C.\ Radin, 
Random close packing of granular matter, 
J.\ Stat.\ Phys.\ 131 (2008) 567-573. 

\item{[Re]} O.\ Reynolds, 
On the dilatancy of media composed of rigid particles
in contact. with experimental illustrations, 
Phil.\ Mag.\ Series 5 20 (1885) 469-481.

\item{[RN]} K.K.\ Rao and P.R.\ Nott,
An Introduction to Granular Flow,
(Cambridge University Press, Cambridge, 2008).

\item{[SNRS]} M. Schr\"oter, S. N\"agle, C. Radin and H.L. Swinney,
Phase transition in a static granular system, 
Europhys.\ Lett.\ 78 (2007) 44004.

\end